\documentclass[onecolumn,prb,color,superscriptaddress,psfig,showpacs,amsmath,amssymb,nobibnotes]{revtex4-1}
\setcitestyle{square,numbers}

\usepackage{amsmath}
\usepackage{graphicx}
\usepackage{graphicx}
\usepackage{float}
\usepackage{epstopdf}
\usepackage{dcolumn}
\usepackage{hyperref}
\hypersetup{
	colorlinks   = true, 
	urlcolor     = blue, 
	linkcolor    = blue, 
	citecolor    = blue 
}
\usepackage{color}
\usepackage{amsmath}
\usepackage{xspace}
\newcommand{\TMX}{(TMTTF)$_2$X }
\newcommand{\DMX}{o-(DMTTF)$_2$X }

\begin{document}

\title{Electron Spin Resonance of Defects in Spin Chains. \DMX : a versatile system behaving like molecular magnet}

\author{L.~Soriano}
\affiliation{Aix-Marseille Universit\'{e}, CNRS, IM2NP UMR 7334, F-13397 Marseille, France}

\author{J.~Zeisner}
\affiliation{Leibniz Institute for Solid State and Materials Research IFW Dresden, D-01069 Dresden, Germany}
\affiliation{Institute for Solid State and Materials Physics, TU Dresden, D-01062 Dresden, Germany}

\author{V.~Kataev}
\affiliation{Leibniz Institute for Solid State and Materials Research IFW Dresden, D-01069 Dresden, Germany}

\author{O.~Pilone}
\affiliation{Aix-Marseille Universit\'{e}, CNRS, IM2NP UMR 7334, F-13397 Marseille, France}

\author{M.~Fourmigu\'{e}}
\affiliation{Universit\'{e} de Rennes, CNRS, ISCR UMR 6226, F-35042 Rennes, France}

\author{O.~Jeannin}
\affiliation{Universit\'{e} de Rennes, CNRS, ISCR UMR 6226, F-35042 Rennes, France}

\author{H.~Vezin}
\affiliation{Universit\'{e} de Lille, CNRS, LASIR UMR 8516, F-59655 Villeneuve d'Ascq, France}

\author{M.~Orio}
\affiliation{Aix-Marseille Universit\'{e}, CNRS, ISM2 UMR 7313, Marseille, France}

\author{S.~Bertaina}
\altaffiliation[]{sylvain.bertaina@im2np.fr}
\affiliation{Aix-Marseille Universit\'{e}, CNRS, IM2NP UMR 7334, F-13397 Marseille, France}
\date{\today}

\begin{abstract}
The paper presents the Electron Paramagnetic Resonance study of defects in the spin chain \DMX family using continuous wave and pulsed techniques. The defects in spin chains are strongly correlated and present similar microscopic structure as a molecular magnet. By means of 2D-HYSCORE and DFT calculations we show a strong reduction of hyperfine coupling between the defects and the nuclear spin bath. We assume that the reduction is due to the Heisenberg exchange interaction which screens the effect of the nuclei.  
\end{abstract}

\maketitle

\section{Introduction}

Molecular magnetic clusters containing a large but finite number of coupled metal ions have been extensively studied in the last decades \cite{Gatteschi2006a,Gatteschi1994,Barra1996a,Wernsdorfer2000b}. Such complex multi-spin systems provide attractive targets for the study of quantum effects at the mesoscopic scale. In general, a molecular magnet refers to a system containing several magnetic ions strongly coupled by exchange interactions which generate a collective spin $S$. The philosophy of such systems is to use the strong correlation to create a gap between the magnetic ground  and  excited states. Among the long list of molecular based magnets we can cite the examples of the Mn12\cite{Gatteschi1994,Park2002a,Chiorescu2000a}, Fe8 \cite{Sorace2003,Park2002a}, V15 \cite{Ajiro2003,Martens2017}...

It is possible to find equivalent structures in S=1/2 one dimensional antiferromagnetic  Heisenberg spin chains (1DQAFM). In uniform 1DQAFM the ground state is magnetic but gapless while in dimerized 1DQADM the ground state is S=0 with a gap with respect to the triplet S=1 state proportional to the exchange coupling constant $J$. In both cases the local magnetization is zero. When a non-magnetic defect is added to the chain, like an end-of-chain, the break in translational symmetry polarizes the spins around the defect \cite{Sorensen1998,Augier1999,Hansen1999,Augier2000,Riera2001,Eggert1992,Eggert1994a,Eggert1995,Eggert1996,Eggert1998,Eggert2001,Fujimoto2005}. 
While open chains in 1DQAFM have been studied in the past by magnetometry \cite{Sirker2007}, and by NMR \cite{Takigawa1997a,Tedoldi1999}  due to the large spin orbit coupling and the difficulty to obtain high enough quality samples in metal oxide, very few EPR studies of end chain has been reported \cite{Smirnov1998}. To reduce the effect of spin-orbit coupling, we have used radical salts rather than metal ions. One of the most famous families of organic magnets, the (TMTTF)$_2X$, with $X$ a counter anion like PF$_6$ or AsF$_6$, also called Fabre salts \cite{Coulon1982} was intensively studied during the last decades and shows an extremely rich phase diagram \cite{Jerome1991,Dressel2012}. In as-grown single crystals, these defects are generally associated with local inhomogeneity such as crystallographic defects or disorder, bond alternations, chain ends, etc. Theoretical predictions for 1DQAFM show that each of these magnetic defects carries an overall spin S = 1/2 originating from by the correlation of dozens of spins \cite{Nishino2000,Nishino2000a}. EPR  measurements on the strongly correlated defects (SCD) in as-grown (TMTTF)$_2$PF$_6$ and (TMTTF)$_2$AsF$_6$ have been reported  \cite{Bertaina2014a}. These systems dimerize below T$_{SP}=19$~K and 14~K for PF$_6$ and AsF$_6$, respectively, with a dimerization parameter $\delta=0.03$ \cite{Dumm2004}. A SCD in such systems consists of about 60 spins and is thus a bit large compared to common molecular magnets. In this paper, we report the electron spin resonance investigation in both continuous mode and pulse mode of SCD induced in o-(DMTTF)$_2$X. We first present a microscopic description of the system and compare it to V15 to show how SCD can be seen as an "induced" molecular magnet. Then, we present the  EPR data of the SCD in \DMX. cw-EPR experiments show how the defects are correlated to the spin chain. A coherence time of the SCD long enough allowed us to perform pulsed EPR which revealed an unexpected reduction of the hyperfine interaction. We propose that the strong reduction of the nuclear bath coupling is due to the many body nature of the SCD and the large exchange coupling.

\section{Microscopic description}
\label{sec:1}

The S=1/2 antiferromagnetic Heisenberg spin chain is the prototype of system in which quantum fluctuations play an important role. Since Bethe \cite{Bethe1931} it is well known that the ground state is not ordered. The spectrum of 1DQAFM is a gapless continuum of states unstable under external interaction. While an isolated quantum spin chain has no N\'{e}el  order even at T=0, interchain interactions open a gap in the ground state leading to the ordered antiferromagnetic state. Interaction with lattice vibration can also lead to the displacement of atoms to minimize the free energy. The exchange interaction is no longer uniform through the chain which dimerizes into S=0 pairs : this is the spin-Peierls state. In this paper we will focus on this last phase. From the theoretical point of view, an isotropic one-dimensional spin system that shows a spin-Peierls transition at low temperatures could be modelled in terms an $S = 1/2$ alternating-exchange Heisenberg chain Hamiltonian \cite{Johnston2000}:

\begin{equation}
\label{eq:HAFM-Hamiltonian}
\mathcal{H} = J\sum_{i}(1-\delta)\boldsymbol{S}_{2i-1}\cdot \boldsymbol{S}_{2i} + (1+\delta)\boldsymbol{S}_{2i}\cdot\boldsymbol{S}_{2i+1} \ \ \ .
\end{equation}

With $\delta$ the explicit dimerization parameter and $J$ the Heisenberg coupling constant. For $\delta=0$ the Hamiltonian describes the uniform Heisenberg chain while for $\delta=1$ it describes an ensemble of isolated dimers.

In \TMX and \DMX families, we have $\delta=0$ for $T>T_{sp}$ and the systems are described by a uniform isotropic Heisenberg spin chain. For $T<T_{sp}$, the chains enter a spin-Peierls phase. This phase is well known and gives rise to a non-magnetic ground state separated from a S=1 triplet. We can now discuss the realistic case of a finite chain (open chain), naturally created during the growing of the crystal when some stacking defects break the translational symmetry. The Hamiltonian becomes:
\begin{equation}
\label{eq:HAFM-HamiltonianFinite}
\mathcal{H} = J\sum_{i=1}^{N-1}(1+p(-1)^i\delta)\boldsymbol{S}_{i}\cdot \boldsymbol{S}_{i+1}  \ \ \ .
\end{equation}

Where $p=\pm 1$ is the parity parameter and $N$ the quantity of spins. The parity parameter describes how the alternation in the interaction of the spin chain start. For $p=+1$ (-1) the chain start with a weak (strong) link. For the sake of clarity, we will not include the next nearest neighbor interactions as well as the elastic coupling with the lattice. A complete treatment of these cases can be found in ref. \cite{Augier1999} . Depending on the defect concentration, $N$ can be very large and becomes mostly unaccessible by exact diagonalization (ED). To describe the finite chain we use the density matrix renormalization group algorithm (DMRG) which can resolve the low-lying states of rather large 1D system (hundreds of spins).
In the following part, the exact diagonalization has been performed by a home-made code while DMRG used the python ALPS toolkit \cite{Bauer2011}.

\subsection{Local Magnetization}

\begin{figure}[h]
	\centering
	\includegraphics[width=0.49\textwidth]{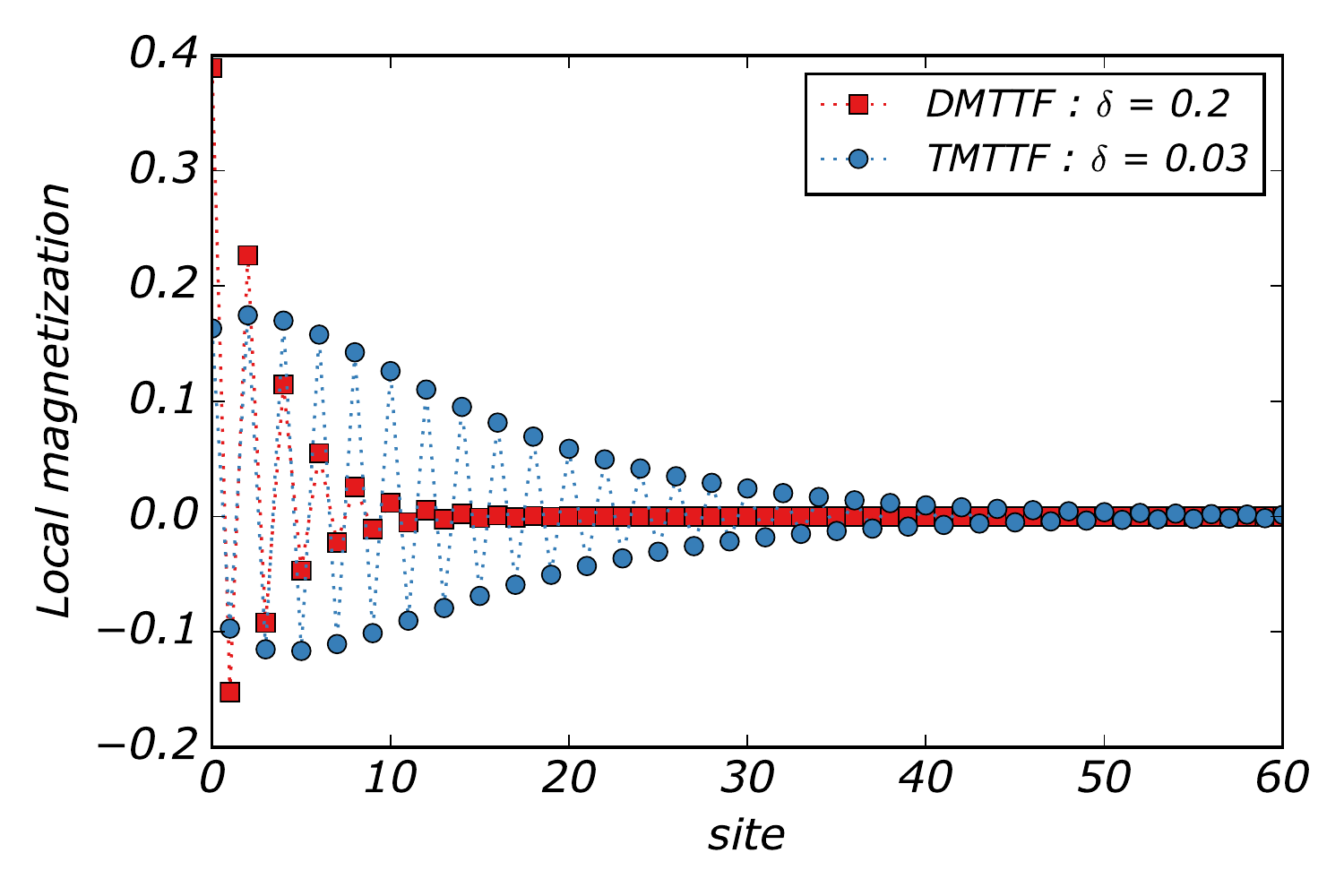}
	\includegraphics[width=0.49\textwidth]{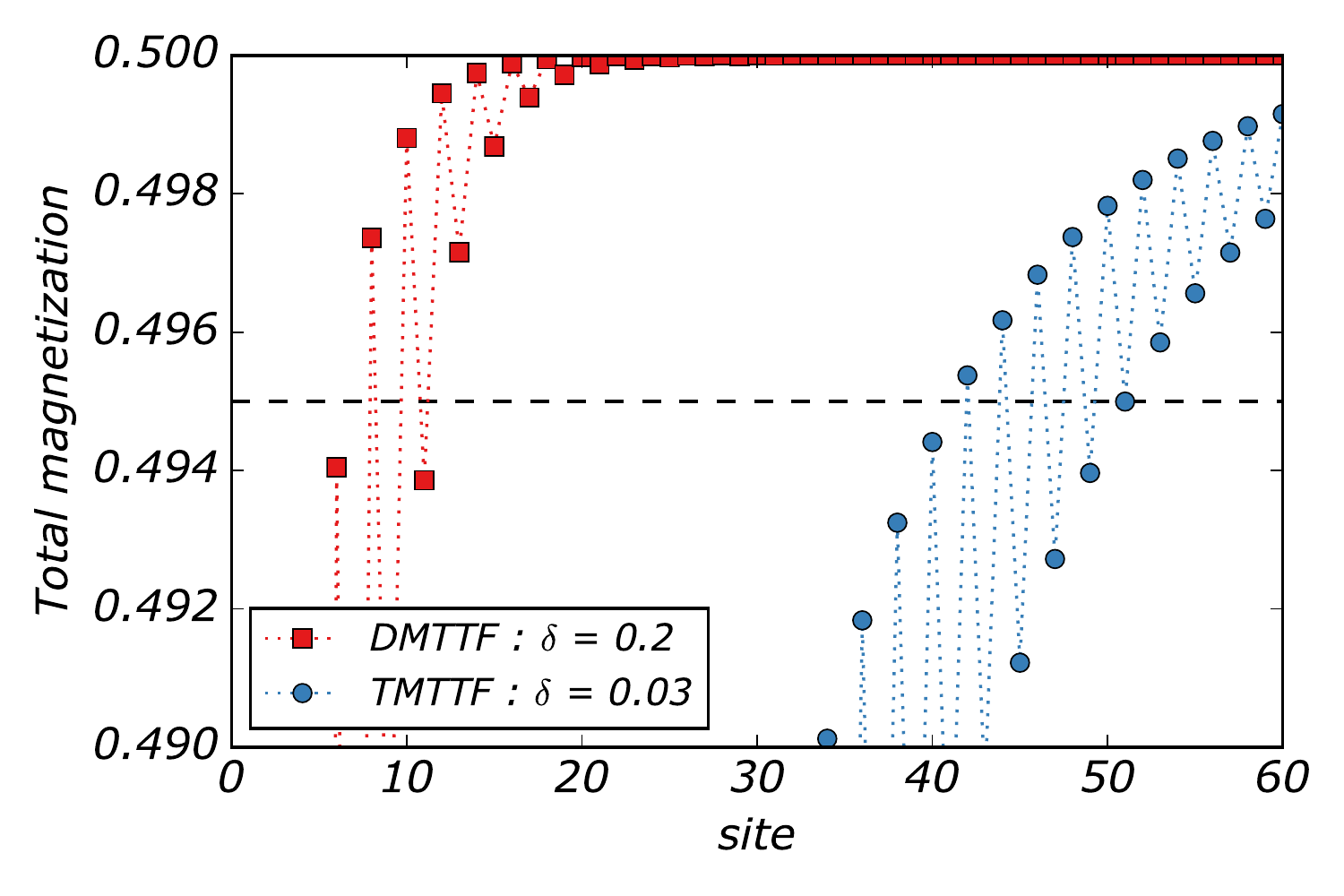}
	\caption{\textit{left} Local magnetization computed by DMRG for 2 explicit dimerization values of $\delta$ with $N=201$ and $p=+1$. The break in translation symmetry at site 0 polarizes many spins around it resulting in a distributed S=1/2 state. Depending on $\delta$, the distribution  polarization remains close to the defect ($\delta$=0.2) or is spread on many sites ($\delta$=0.03). \textit{right} Cumulative sum of the local magnetization for \DMX and \TMX. The dashed line is the limit where 99\% of the magnetization is localized     }
	\label{Fig:LocMag}
\end{figure}

We calculate eq. \eqref{eq:HAFM-HamiltonianFinite} for $N$ up to 201. Using DMRG we then extract the energy, $S_z$, and local magnetization of the ground state and the first excited state as function of the number of spin $N$, the parity $p$ and the dimerization factor $\delta$. We can distinguish two cases: (i) for even values of $N$ the ground state is non magnetic ($S_z=0$) with a large gap $\Delta$ for $p=-1$ and a small gap for $p=1$. While in the case (ii) of odd value of $N$ the ground state is $S_z=1/2$ with no effect of the $p$ in the energy spectrum. 
The local magnetization  of $N=61$ and $p=1$ is presented in Fig.\ref{Fig:LocMag}. We choose two values of $\delta$ corresponding to two kind of organic spin chains: \TMX has a weak dimerization factor $\delta=0.03$ while \DMX has a strong factor $\delta=0.2$ \cite{Jeannin2018}. We can observe an important effect of the translational symmetry breaking at site 0. While in an infinite spin-Peierls chain, one can expect a non magnetic ground state, here the defect polarizes many spins around it. It is important to notice that the total magnetization remains $S_z=1/2$ whatever the values of $\delta$ and $N$ are. For $p=-1$ the local magnetization structure is the same but the polarized cluster is located on the right side of the chain rather than on the left side. The main difference between \TMX and \DMX is the amount of spins needed to contain the total magnetization. Fig. \ref{Fig:LocMag} (right  part) shows the cumulative sum of the local magnetization. Starting from 0 (not shown) it increases to reach 1/2. The number of sites needed to reach 99\% of the full magnetization depends on $\delta$.  In the case of \DMX ($\delta=0.2$), one needs about 13 spins to carry 99\% of the magnetization while in \TMX, one needs about 50 spins. Note that for $\delta$=0 (uniform chain), one needs all the sites of the chain to carry the polarization.             

\subsection{Gap evolution}
\begin{figure}[h]
	\centering
	\includegraphics[width=0.49\textwidth]{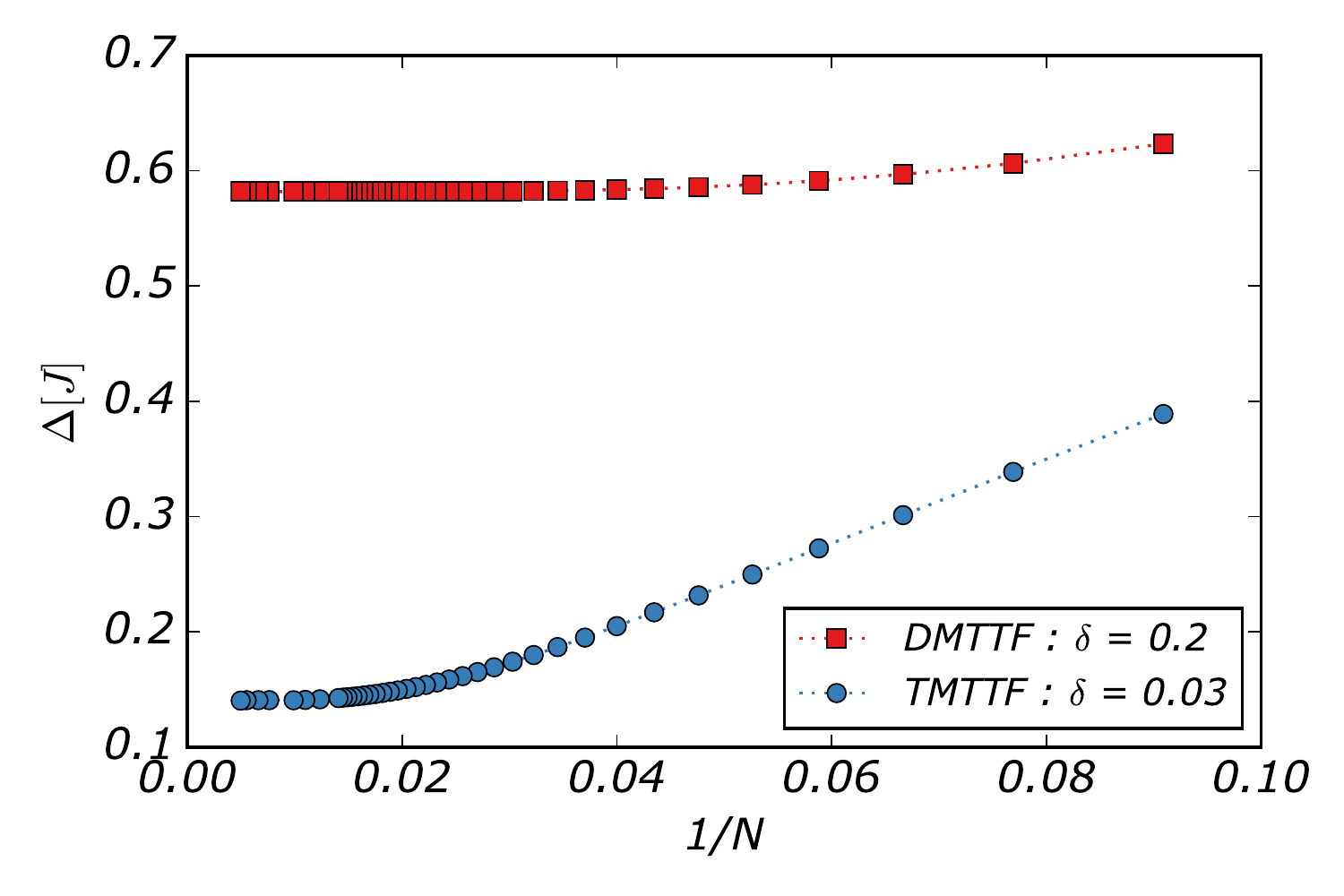}
	\includegraphics[width=0.49\textwidth]{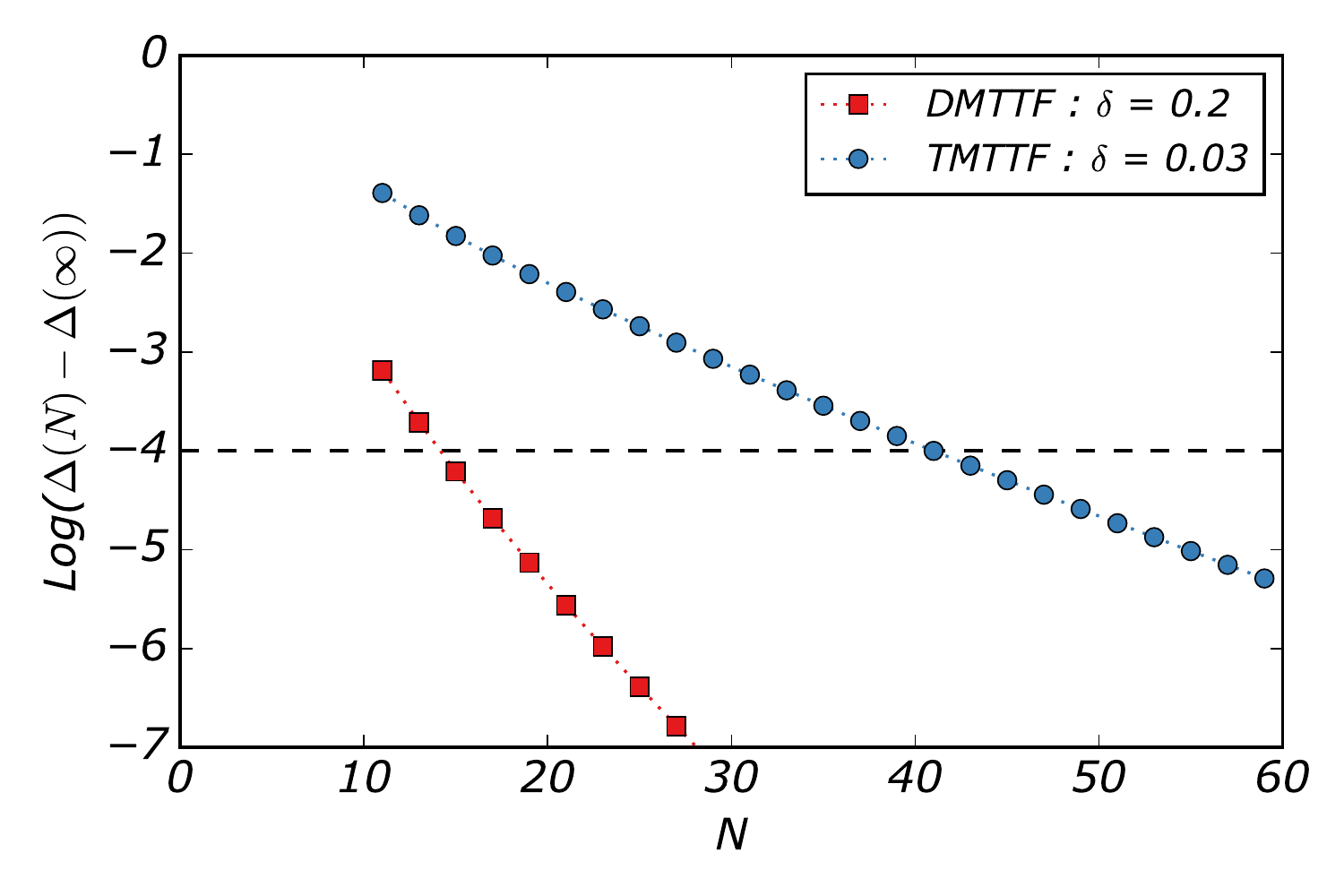}
	\caption{\textit{left}  Chain size dependence of the gap between the ground state and the first excited state for $\delta$=0.2 and 0.03.\textit{right} Difference between the gap $\Delta(N)$ for a chain of length $N$ and $\Delta(\inf)$ for an infinite chain. The dashed line corresponds to a difference $<10^{-4}$.   }
	\label{Fig:Gap}
\end{figure}

We now discuss the effect of $N$ and $\delta$ on the gap between the ground state and the first excited state. Fig.\ref{Fig:Gap} shows the evolution of the gap with $N$. When $N$ increases the gap decreases until it reaches a threshold value independent of $N$.
One can see that, for \TMX, at least 40 spins are needed to get 99.99\% of the gap of the infinite chain while only 13 are required for \DMX. In other words, even if the chain has hundreds of spins, only a few of them are required to describe the low-energy states.  

\subsection{Exact diagonalization}
\begin{figure}[h]
	\centering
	\includegraphics[width=0.49\textwidth]{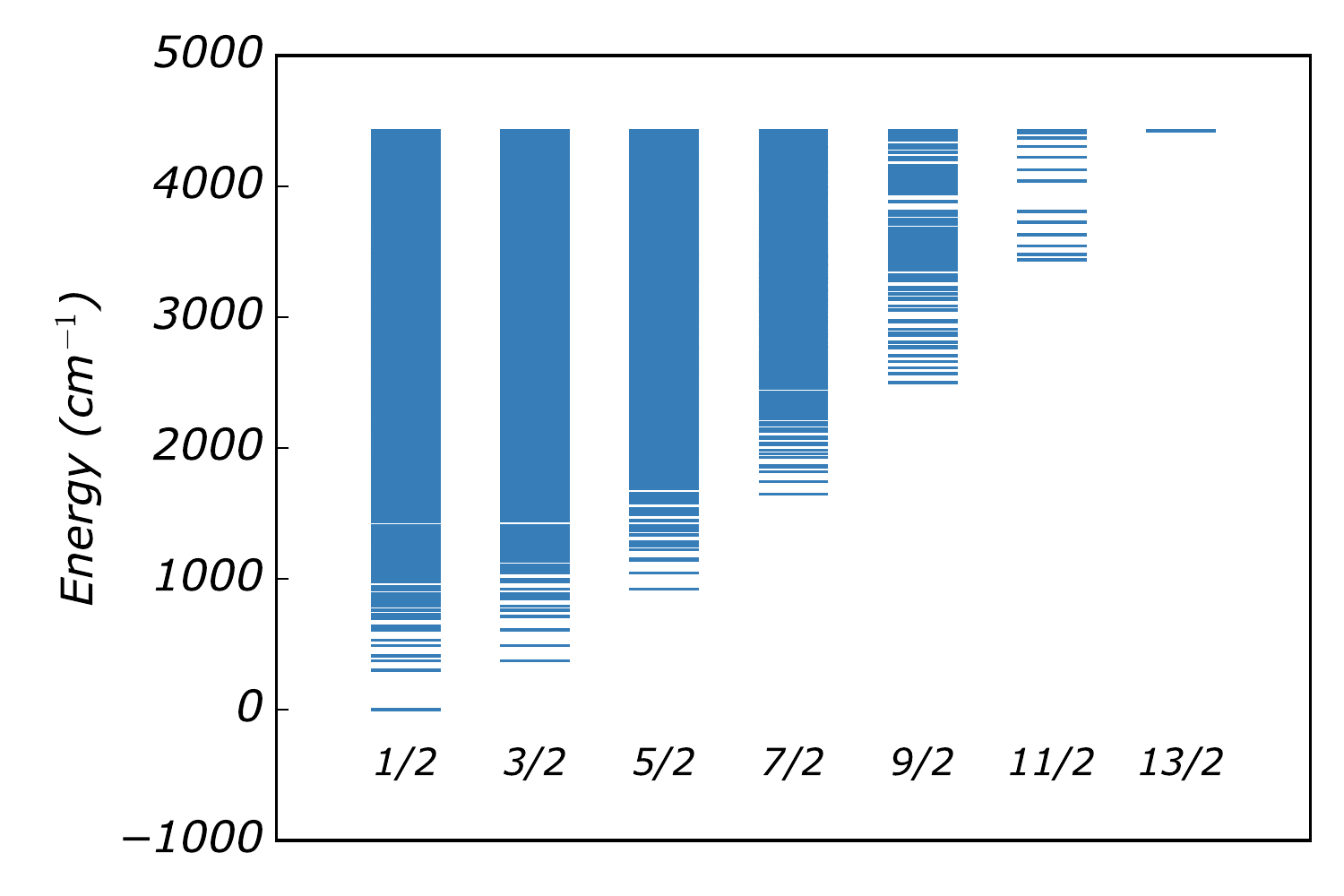}
	\includegraphics[width=0.45\textwidth]{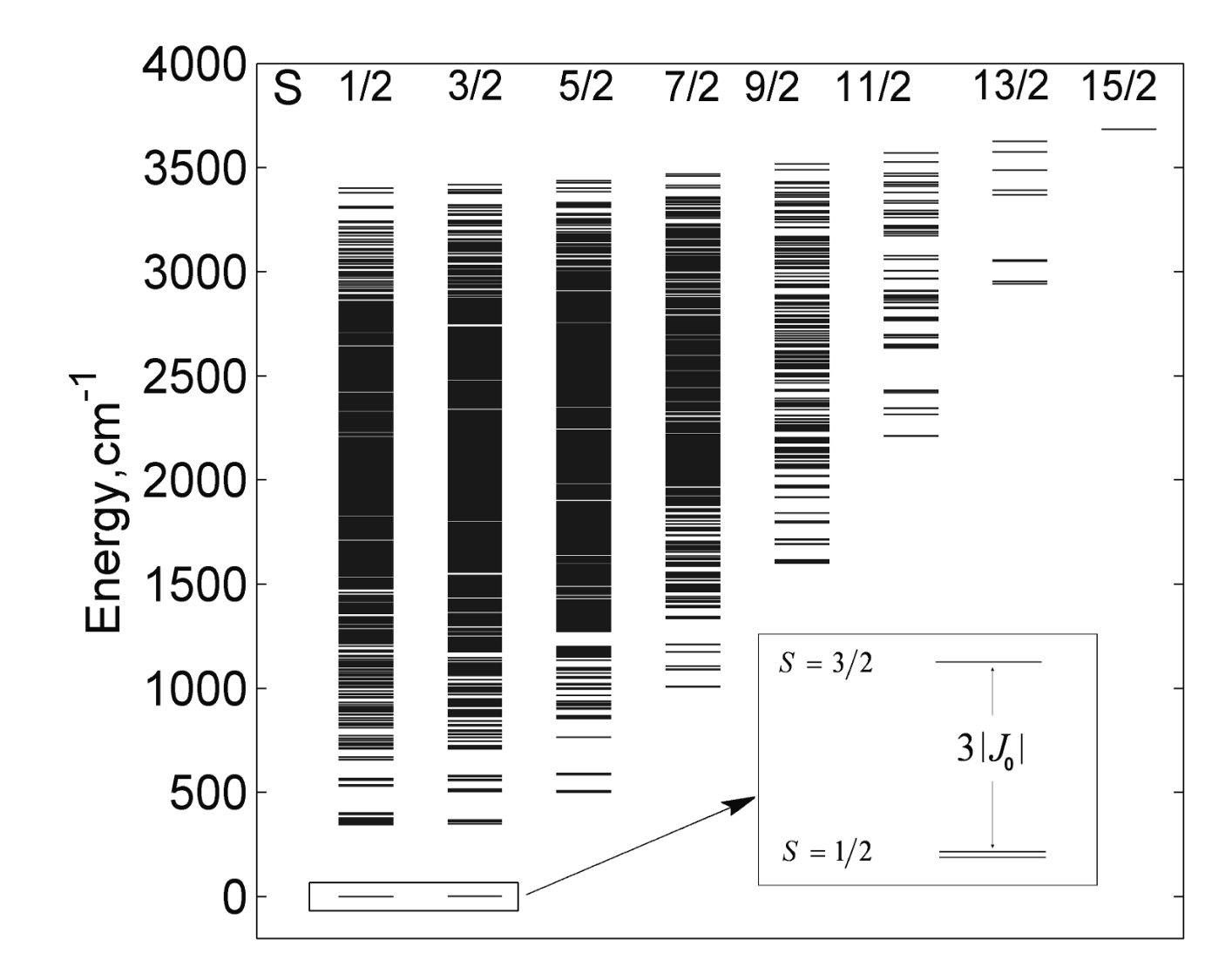}
	\caption{\textit{left}  Exact diagonalization of eq.\eqref{eq:HAFM-HamiltonianFinite}  for $N=13$ and $\delta$=0.2. In absence of strong anisotropy $S_z$ is good quantum number and we can plot the energies by sectors. \textit{right}  Exact diagonalization of V15 adapted with permission from \cite{Tsukerblat2006a} .  }
	\label{Fig:Exact}
\end{figure}

The previous results show that the magnetic structure of a defect in a large spin chain needs only a few number of spins to be accurately described. In \TMX, this number remains too large to be computed numerically while in \DMX only 13 to 15 spins are necessary. Without loss of generality, we expect that the results from \DMX can be extrapolated to \TMX. Fig.\ref{Fig:Exact} shows the result for the exact diagonalization of eq.\eqref{eq:HAFM-HamiltonianFinite} for $N=13$. The $2^{13}=8192$ states are separated in $S_z$ sectors. One can see that the ground state is $S_z=1/2$ separated by a gap to several excited states $S_z=1/2$ and $S_z=3/2$ and then a quasi-continuum of states.   We first compare the energy spectrum of a SCD to the one of a molecular magnet. Fig. \ref{Fig:Exact} (right) shows the energy of V15 \cite{Tsukerblat2006a}. Here the ground states $S_z=1/2$ and $S_z=3/2$   are almost degenerate but the next excited states are separated by a large gap directly proportional to the exchange interaction between the vanadium ions. 

\section{Experimental Details.}

\subsection{Fabrication of \DMX}

\begin{figure}[h!]
	\includegraphics[width=0.49\textwidth]{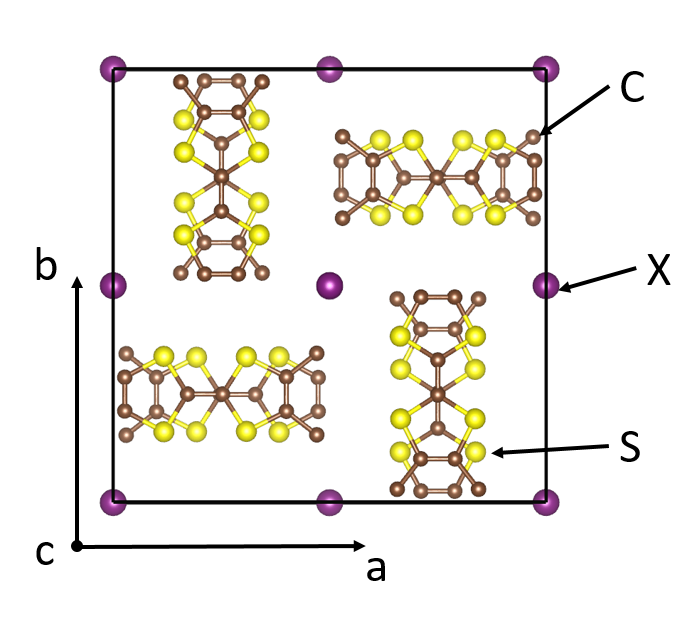}
	\includegraphics[width=0.49\textwidth]{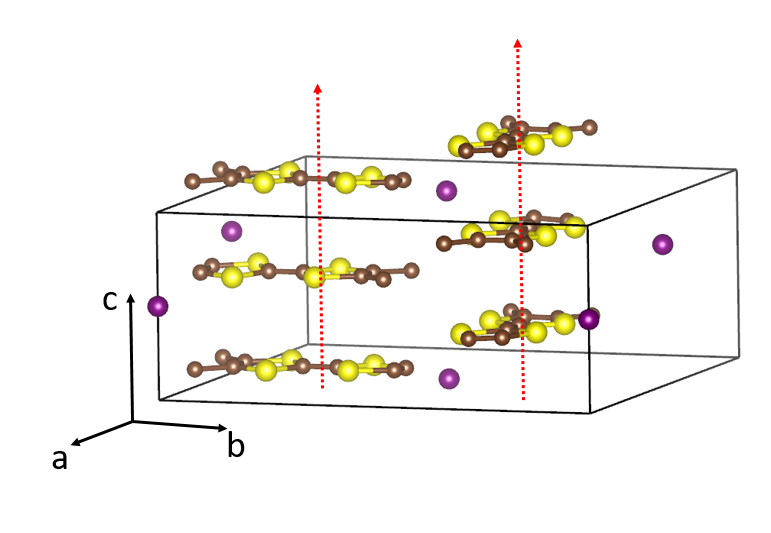}
	\caption{Crystallographic structure of \DMX. The unit cell contains 4 molecular unit orthogonal from each other. The staking axis $c$  is represented by red arrows. The Counter anion $X$ is I, Br or Cl and does not dramatically change the structure (see SI) }
	\label{Fig:structure}
\end{figure}

Crystals of \DMX with X= Br, I and Cl were synthesized by electrocrystallization methods as described in \cite{Fourmigue2008}. Their structural, electronic and basic magnetic characterization was reported in \cite{Fourmigue2008,Foury-Leylekian2011}. The crystallographic structure is shown in Fig.~\ref{Fig:structure}. The tetragonal unit cell (space group $I\overline{4}2d$ (no. 122) \cite{Fourmigue2008}) comprises eight planar DMTTF molecules which form stacks along the crystallographic $c$ axis and which are rotated by 90$^{\circ}$ with respect to the neighboring stacks in the $ab$ plane. As a consequence of this rotation, interactions between neighboring stacks are weak, leading to one-dimensional electronic properties \cite{Fourmigue2008,Foury-Leylekian2011}. Each DMTTF molecule gives one electron to the counter-anion X so that the hole (and the spin) is shared by two DMTTF. Like \TMX, the \DMX family presents a rich phase diagram \cite{Foury-Leylekian2011}. Conductor at room temperature, it exhibits a charge localisation below $T_{CO}$ then below $T_{SP}\sim50$~K the (DMTTF)$_2$ stacks dimerize in the direction of the chain $c$ inducing the spin-Peirls state.  Above   $T_{SP}$ the spin system can be described by an $S = 1/2$ Heisenberg AFM uniform chain model in which each double molecule of a stack corresponds to one site in the spin-chain lattice which is occupied by a single spin associated with the localized hole. Below $T_{SP}$ the systems are described by the explicit dimerized spin chain \eqref{eq:HAFM-Hamiltonian}. 

The crystals grow in a needle-like shape with the longest side of the crystal corresponding to the $c$ axis (the chain axis). Thus, the tetragonal symmetry and the shape of the crystals make them easy to orientate in a magnetic field. The crystal is rather fragile and sensitive to thermal dilatation. Although we took a particular attention when  handling the crystal and controlling the speed of the temperature ramp, we used a fresh sample for each set of measurements. The defects, which are the central part of this article, are naturally induced during the crystallization and  can then slightly change between batches.    

\subsection{Electron Paramagnetic Resonance}

Continuous wave (CW) EPR measurements were performed using Bruker EMX X-band spectrometer operating at microwave (mw) frequency of about 9.6\,GHz and in fields up to 1.3~T. This spectrometer is equipped with standard He-flow cryostats which allow measurements in the temperature range from 4~K up to room temperature. Moreover, a goniometer installed at the spectrometer enable measurements of EPR parameters as a function of the angle between the crystal axes and the external magnetic field. For pulse EPR measurements at X-band frequency ($\sim$9.7\,GHz) a Bruker Elexsys E580 spectrometer was used. It is equipped with a cryogen-free cryostat. Relative orientation of the chain axis $c$ and the magnetic field is insured by a goniometer with an angle $\theta=0^\circ$ corresponding to $H\parallel c$

We use a combination of high-resolution two-dimensional (2D) hyperfine sublevel correlation (HYSCORE) spectroscopy and density functional theory (DFT) calculations to determine the electron-nuclei interaction. Contrary to standard HYSCORE study, we didn't focus on a specific nuclear spin but we probed the correlation with all the nuclear spins present in \DMX.  
HYSCORE experiments performed at a frequency of 9.728~GHz. The echo amplitude was measured using the 2D-HYSCORE pulse sequence ($\pi/2-\tau–\pi/2–t_1–\pi–t_2–\pi/2–\tau$–echo) with 16 ns and 32 ns for the $\pi/2$ and $\pi$ pulses, respectively. $t_1$ and $t_2$ are increased by 24ns over 256x256 steps. The unwanted echoes were eliminated by applying a 16-step phase cycling procedure. The choice of $\tau$ is important since it is responsible of blind spots. Since we wanted to observe all nuclear spin effects, the sequence was performed at many values of $\tau$ (300ns, 350ns, 450ns, 650ns). 

\subsection{DFT calculation details}

All theoretical calculations  based on the Density Functional Theory (DFT), were performed with the ORCA program package \cite{Neese2012}. To facilitate comparisons between theory and experiments, X-ray crystal structure of o-(DMTTF)$_2$I was used. Our DFT molecular model was built considering two dimethetyltetrathiafulvalene units together with 8 iodine counter-ions. This model was then optimized while constraining the positions of all heavy atoms to their experimentally derived coordinates. Only the positions of the hydrogen atoms were relaxed because these are not reliably determined from the X-ray structure. Geometry optimization as well as electronic structure calculations were undertaken using the hybrid functional B3LYP \cite{Becke1993,Lee1988} in combination with the TZV/P\cite{Schafer1994} basis set for all atoms, and by taking advantage of the resolution of the identity (RI) approximation in the Split-RI-J \cite{Weigend2006} variant with the appropriate Coulomb fitting sets \cite{Klamt1993}. Increased integration grids (Grid4 and GridX4 in ORCA convention) and tight SCF convergence criteria were used in the calculations. EPR parameters including g-tensors and hyperfine coupling constants were obtained from relativistic single-point calculations using the B3LYP functional. Scalar relativistic effects were included with ZORA paired with the SARC def2-TZVP(-f) basis sets \cite{Pantazis2008,Pantazis2009} and the decontracted def2-TZVP/J Coulomb fitting basis sets for all atoms. The integration grids were increased to an accuracy of 5 (ORCA convention) and the picture change effects were applied for the calculation of the hyperfine tensors. All chemical structure images were generated using the orca\_plot utility program and visualized with the Chemcraft program.

\section{Results and discussions}
\subsection{cw-EPR spectroscopy}

\begin{figure}[h]
	\centering
	\includegraphics[width=0.49\textwidth]{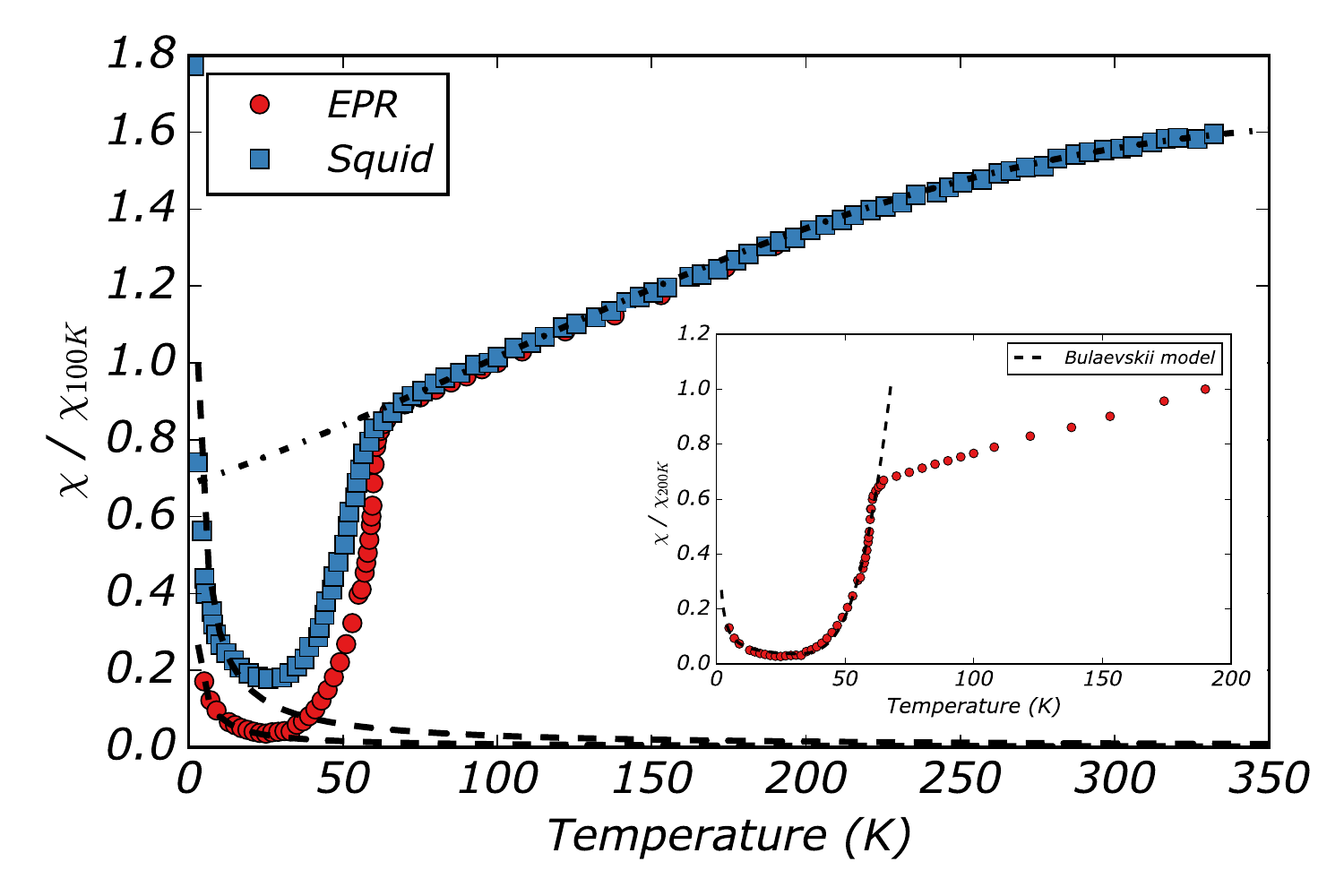}
	\includegraphics[width=0.49\textwidth]{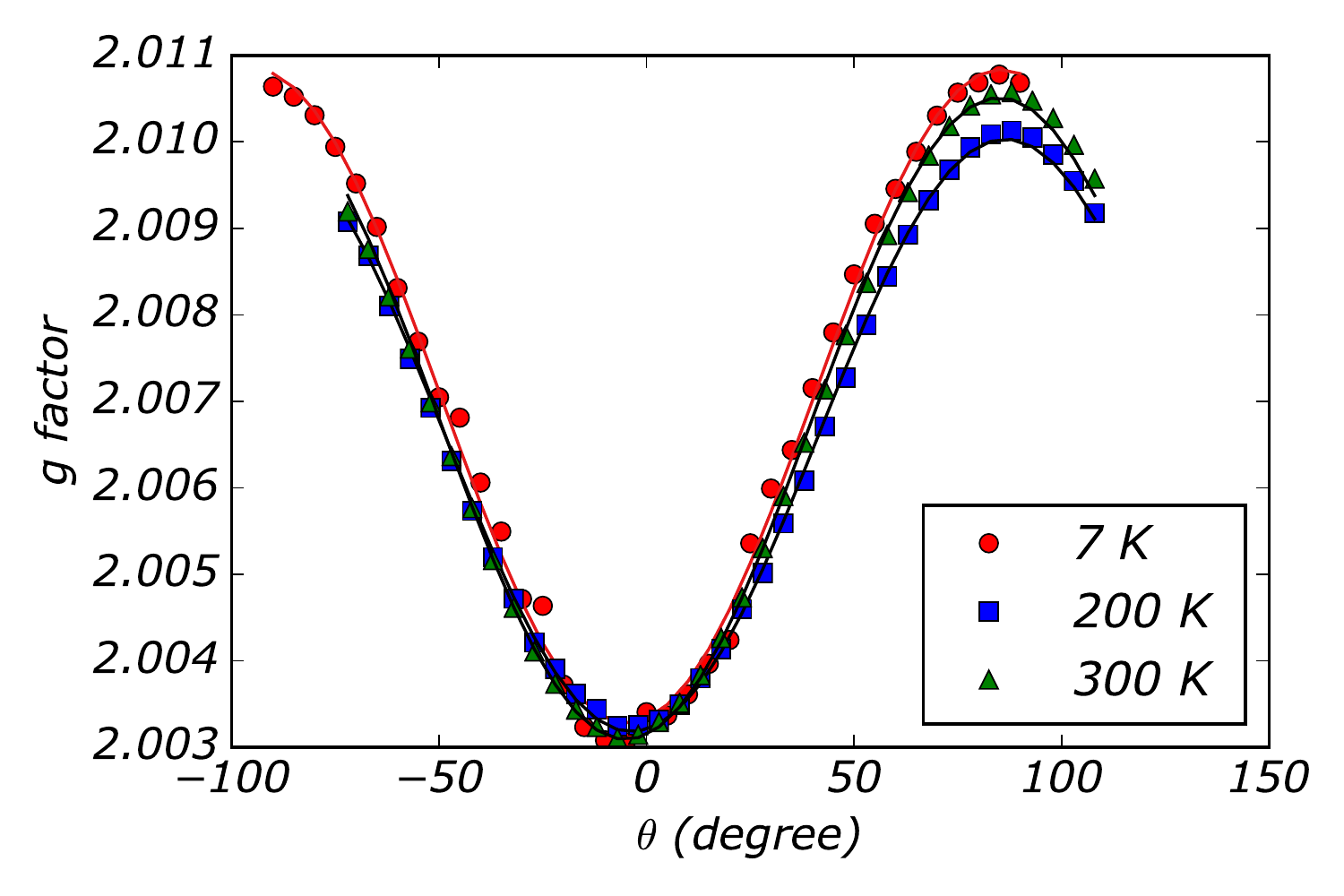}
	\caption{\textit{left}  Temperature dependence of the EPR susceptibility $\chi_{EPR}$ (this study) and DC susceptibility $\chi_{DC}$(from \cite{Foury-Leylekian2011}) of o-(DMTTF)$_2$I. To compare them, $\chi_{EPR}$ and $\chi_{DC}$ are renormalized by their value at $T=$100~K . For $T>60$~K, $\chi_{EPR}$ and $\chi_{DC}$ are fitted using Eq. \eqref{eq:BF} (dashed-dot line). For $T<25$~K the data are fitted using the Curie-law (dashed lines) The inset shows the best fit using the Bulaevskii model \cite{Bulaevskii1969} .  \textit{right} Angular dependence of the g-factor for three temperatures.   }
	\label{Fig:EPR1}
\end{figure}

In the following, we present  CW-EPR data of (DMTTF)$_2$I. The results on Br and Cl counter-anions have been reported previously \cite{Foury-Leylekian2011,Zeisner2019a}. Due to the conductive nature of \DMX at high temperature, the EPR line is a dysonian at room temperature and when the temperature decreases, it progressively becomes a pure Lorentzian \cite{Coulon2007}. At low temperature ($T<10$~K), the line is a mixing between a Gaussian and a Lorentzian shape. From the EPR line, we can extract the intensity, the g-factor, the linewidth (not presented) and the dispersion (not presented).      
Fig. \ref{Fig:EPR1} shows the temperature dependence of the EPR intensity (left) and the angular dependence of the g-factor (right). Thanks to the Kramers-Kronig relation the intensity of the EPR line ($\chi_{EPR}$) is directly proportional to the dc-susceptibility   $\chi_0$. For $T>T_{SP}=$70~K the system is described by a uniform Heisenberg spin chain and its susceptibility can be fitted by the Bonner-Fisher (BF) \cite{Bonner1964} formula: 
\begin{equation}\label{eq:BF}
\chi = \frac{N(g\mu_B)^2}{k_BT}\frac{0.25+0.074975x+0.075235x^2}{1+0.9931x+0.172135x^2+0.757825x^3}
\end{equation}
With $x = \frac{J}{k_BT}$.
The best fit of the BF equation gives access to the exchange coupling constant $J=650$~K. This value has to be taken with caution since no temperature variation of the volume was taken into account like in \TMX  \cite{deSouza2013,Salameh2011}. 

Below $T_{SP}$ the system dimerized while entering in the spin-Peierls phase \cite{Fourmigue2008} and $\chi_{EPR}$ quickly drop. Below $T=$25~K the susceptibility rise again with a Curie like behavior.

We will now focus on this last Curie-tail. It is common in SQUID magnetometry to observe such a behavior which originates from paramagnetic impurities. Most of the time, it's treated as extrinsic to the magnetic object under  study and is just remove from the final result. The drawback of dc-magnetometry is all magnetic contributions are summed into one point while EPR add a spectral dimension to this point so it is possible in principle to separate the contributions. Fig. \ref{Fig:EPR1} (right) shows the angular dependence of g-factor at temperatures where only the uniform chain gives a signal (T=200K and 300K) and at temperature where the spin chain is dimerized (T=7K) and so only the impurity can give a signal. For the whole range of temperature, both the values and the anisotropy of g-factor are the same while one can expect a different g-factor (certainly isotropic) of an extrinsic impurity.  In the case of extrinsic impurities, the g-factor is expected to be isotropic or at least different than the spin chain. The EPR signal of defects in spin chains has previously been reported in \TMX \cite{Bertaina2014a} and \DMX \cite{Zeisner2019a}  but very few data were collected for metal ion systems like CuGeO3 \cite{Smirnov1998} where the spin orbit coupling makes the line too broad to be easily observed. Note that the absolute value of g-factor is not so accurate since a small field remanance in the electromagnet can shift the value of the real g-factor. However the relative value $\Delta g= g(90^\circ)-g(0^\circ)=8.10^{-3}$ is given with a good precision.
 
The quantity of SCD can be estimated using the following procedure and consideration:
We assume the Curie tail to be described by a the Curie law such as:
 \begin{equation}
\label{eq:curie_law}
\chi_{LT} = \frac{g^2\mu_B^2S(S+1)N'}{3k_BT}
\end{equation} 
Where $N'$ is the number of SCD. This might be an important approximation since it is not determined whether defects in spin chain needs a low temperature normalization like in uniform chain  \cite{Eggert1994a}. However Eq.\eqref{eq:curie_law} was used to determine the number of defects in CuGeO3 \cite{Grenier1998}. The problem is that EPR is not a direct quantitative technique and one needs to compare it with something else. In our case, since we can describe the uniform chain  by the Bonner-Fisher curve we can use the relation:
\begin{equation}
\frac{J \chi_{max}}{g^2\mu_B^2N} \simeq 0,1469
\label{eq:chi_max}
\end{equation} 
Where $\chi_{max}$ is the maximum of the susceptibility of the spin chain and $N$ the number of spin in the chain. 
Combining Eq.\eqref{eq:curie_law} and Eq.\eqref{eq:chi_max} we get:

\begin{equation}
\label{eq:defects_concentration}
\frac{N'}{N}= d = \frac{0.1469 \times 4C}{J\chi_{max}} 
\end{equation}
With $J\sim 550$~K, $\chi_{max}$ and $C$  obtained from the fit procedure, we extract a density of defects measured by EPR of $d=5.2.10^{-4}$ while  by SQUID the density of overall impurities is  $d=2.10^{-3}$. 

The density of SCD probed by EPR gives us also the average distance between 2 SCD which in 1D magnetic system is simply \cite{Sirker2007} $\bar{L}=1/p-1\simeq 2000$ sites.

\subsection{pulsed EPR spectroscopy}
In the following we present the 2D-HYSCORE experiments performed on single crystals of \DMX. The SCD is used to  probe the hyperfine interaction with the nuclear spin of \DMX. 

\begin{figure}[h]
	\centering
	\includegraphics[width=0.49\textwidth]{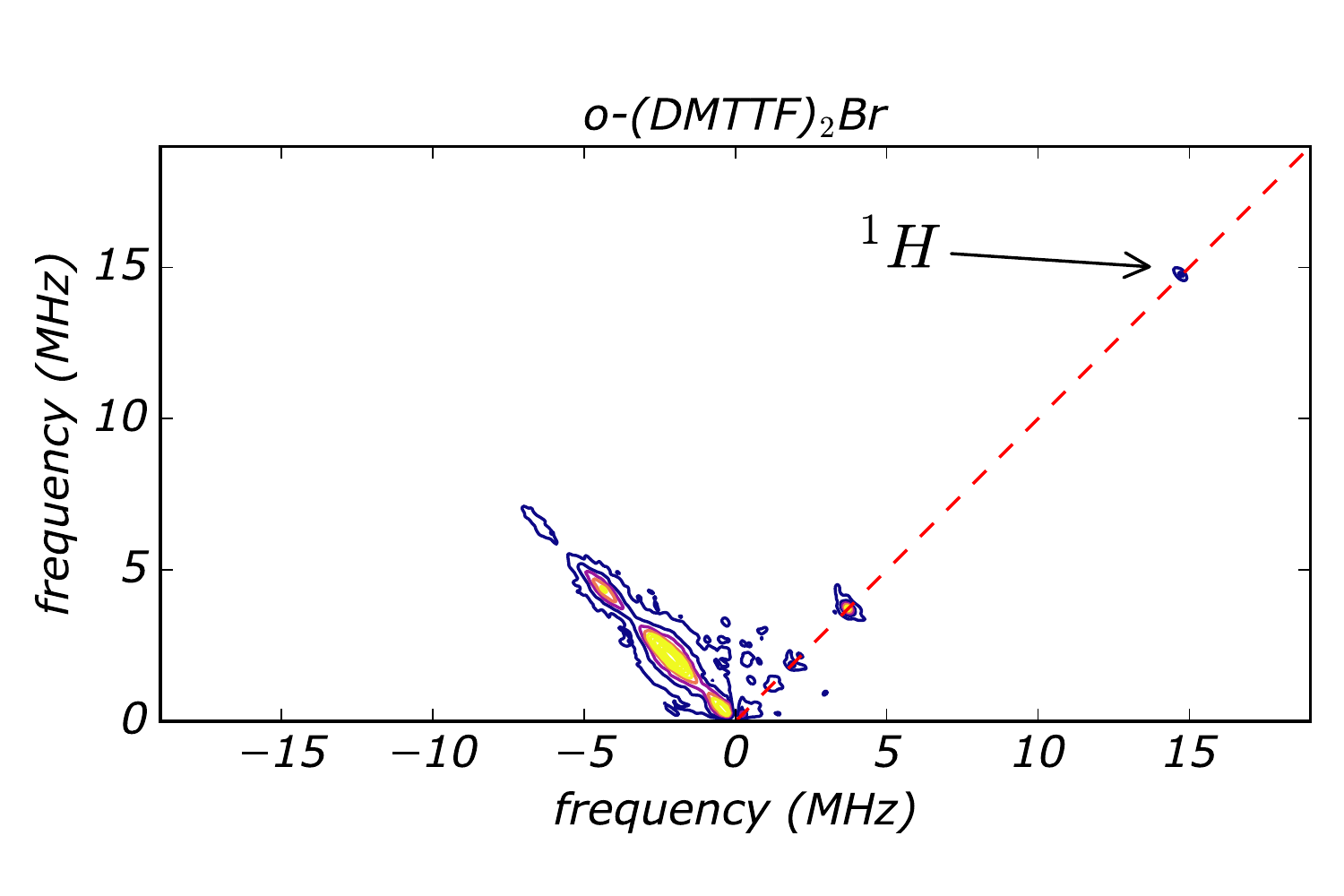}
	\includegraphics[width=0.49\textwidth]{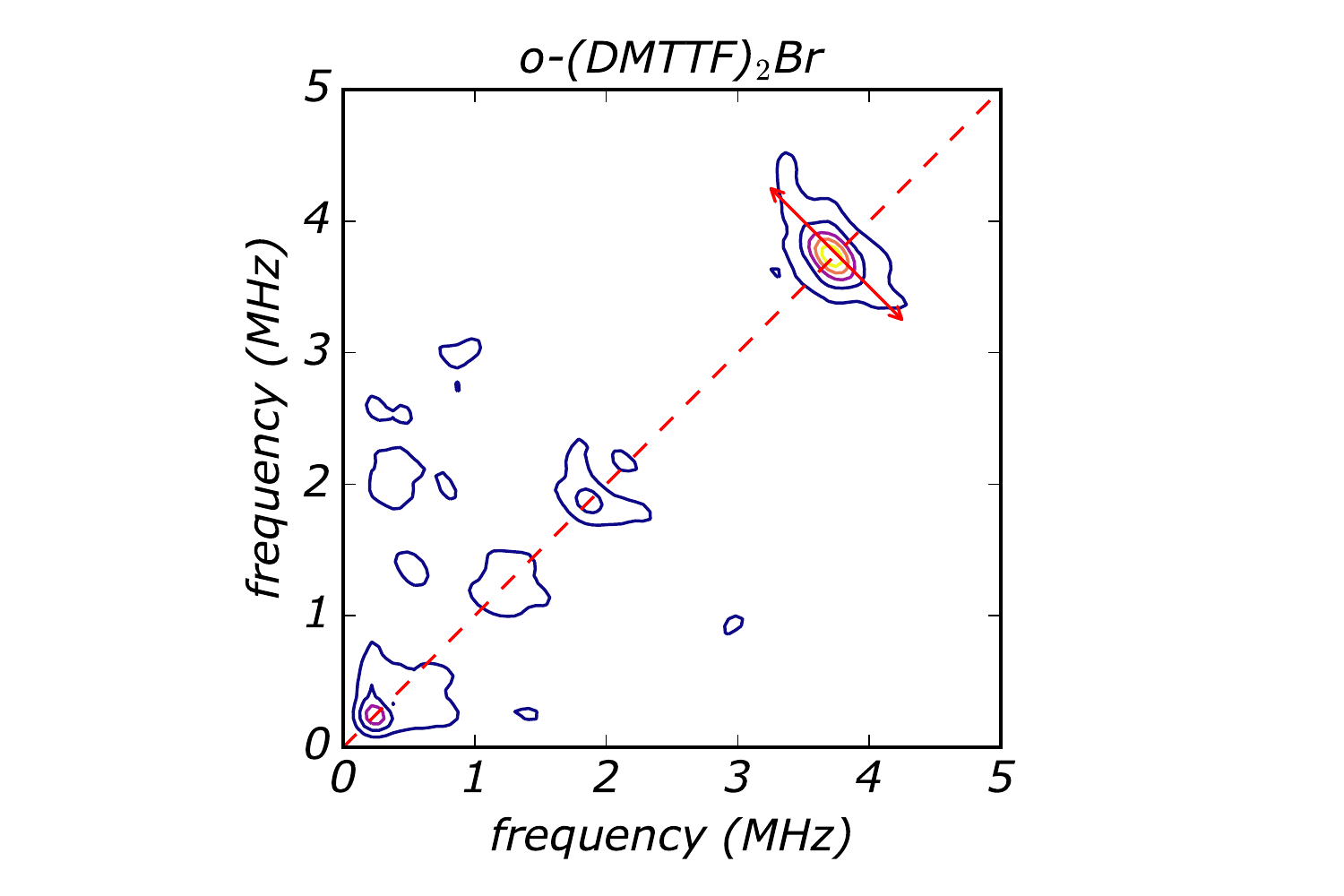}
		\includegraphics[width=0.49\textwidth]{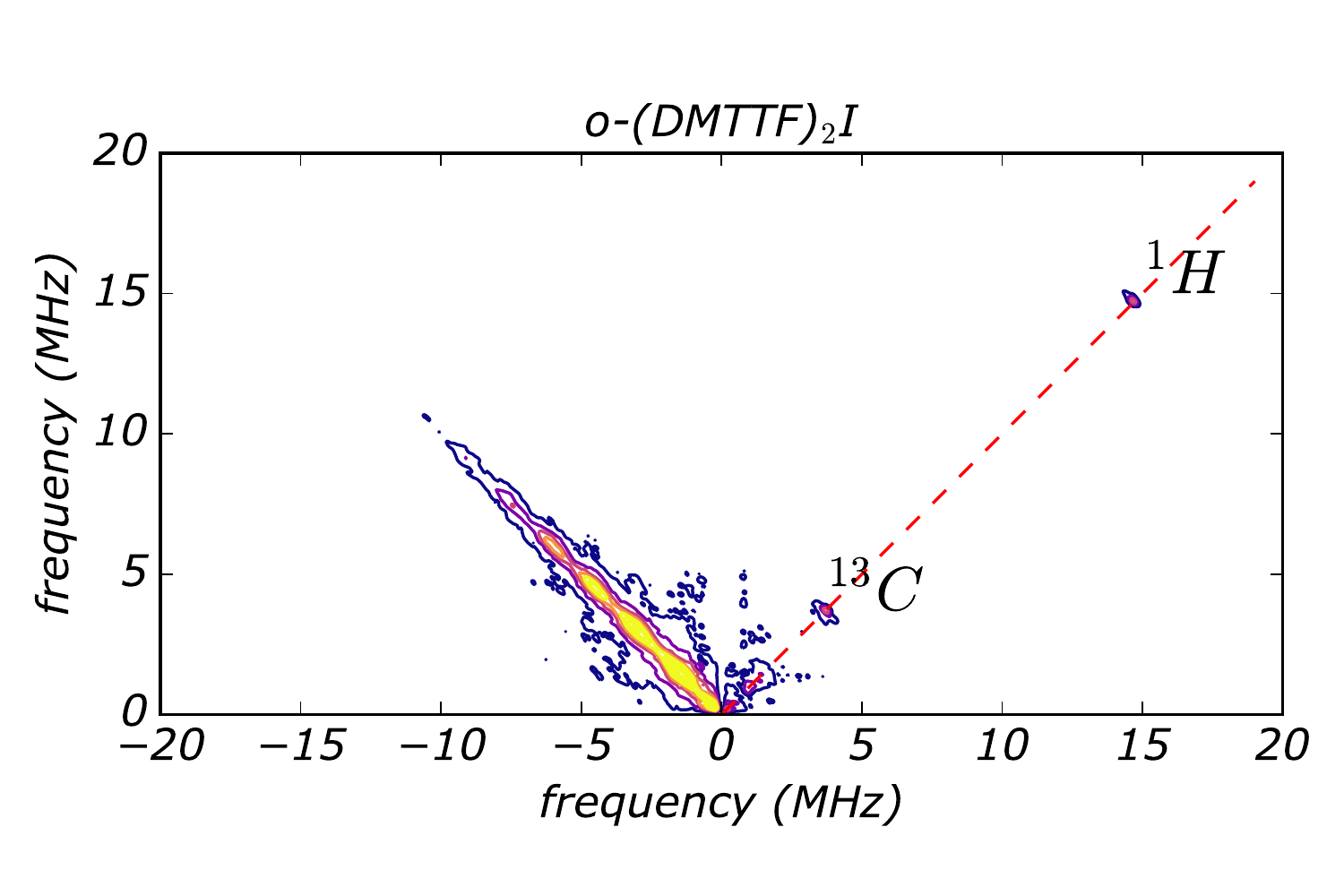}
	\includegraphics[width=0.49\textwidth]{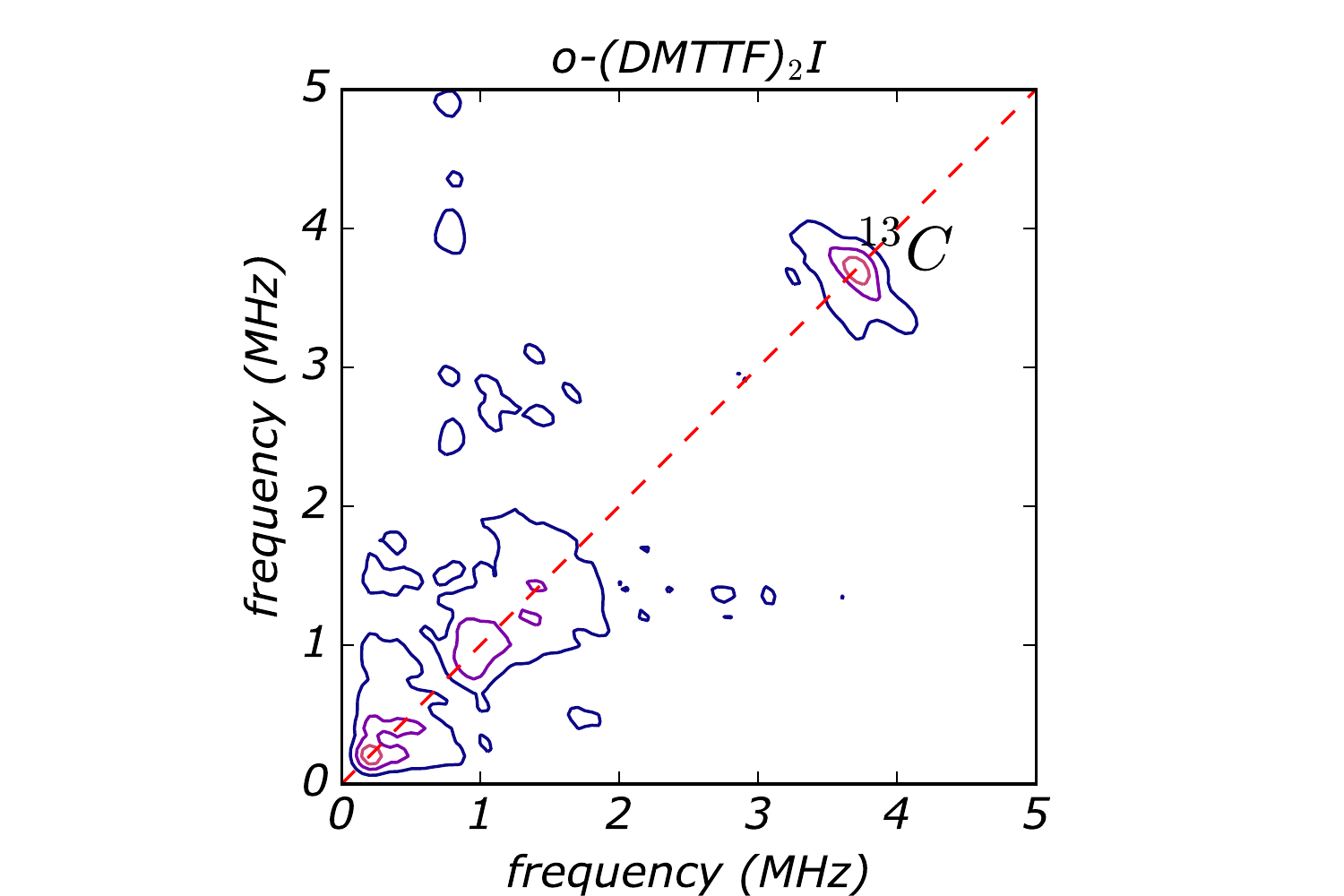}
	\caption{2D-HYSCORE of (DMTTF)$_2$Br (\textit{top}) and  (DMTTF)$_2$I   (\textit{bottom}) recorded at T=6K for $H\perp c$       The right figures are a closer view in the range of 5~MHz}
	\label{Fig:Hyscore}
\end{figure}

Fig. \ref{Fig:Hyscore} shows the HYSCORE contour plot of (DMTTF)$_2$Br and  (DMTTF)$_2$I recorded at T=6~K for $H$ perpendicular to the chain axis. One can see some spots in the positive quadrant (weak coupling) mostly on the diagonal. The negative quadrant exclusively shows intense antiphase echoes  on the diagonal which is usually due to imperfect $\pi$ pulses. 

Changing the crystal orientation, the $\tau$ value or the pulse length did not drastically change the result. Fig. \ref{Fig:Hyscore} shows optimized sequence used to resolve the maximum number of  peaks. Intriguingly, the 2D-HYSCORE data are rather simple and can be described graphically avoiding any complicated simulation. We observe the characteristic spot of the proton $^1$H at 14.7~MHz at the field used in X-band. A closer look at low frequencies shows a few more spots identified at 3.7~MHz, 1.9~MHz and 1.2~MHz. Some signal exist below 1MHz but are unresolved and will need a high frequency study. Regarding (DMTTF)$_2$Br, the spot at 3.7~MHz can be attributed to both the $^{13}$C (abundance 1.07\% )and $^{79}$Br (abundance 50.7\%). However $^{81}$Br at 3.9MHz (abundance 49.3\%)) is not observed and the spot at 3.7MHz is also present in (DMTTF)$_2$I where no Br is involved.   $^{127}$I (abundance 100\%) at 2.96~MHz was not observed. We conclude that the SCD doesn't see the counter anion but interact with $^{13}$C. Regarding the spot at 1.2~MHz, This frequency is close to the Larmor frequency of  $^{33}$S (1.14~MHz) but in the limit of resolution.  The natural abundance is very low (0.75\%) but since we clearly observe $^{13}$C nuclei, the detection of $^{33}$S seems reasonable. However, we have to be aware of the possibility to detect  $^{14}$N which is highly abundant. The Larmor frequency is close to  $^{33}$S (see SI). It is not present in \DMX but is a part of the reagents used during the synthesis  of the crystals. Residues of complexes containing nitrogen are repelled at the surface of the crystal during the growing process and a particular attention has been devoted to rinse the crystals and remove the maximum of residues. 

The frequency size of the spots can give us some information on the hyperfine coupling constant $A$, at least on the upper threshold. From Fig. \ref{Fig:Hyscore} one obtains for $^1$H  $A_H<$0.7MHz, for $^{13}$C, $A_C<$1.3MHz and for $^{33}$S, $A_S<$0.4MHz. Compared to uncorrelated organic radical magnets, these values are very low.

\subsection{DFT calculation}

The hyperfine coupling constants found experimentally are rather small. We think that these small values are the consequence of the exchange coupling in the strongly correlated defects. The large number of elementary cells involved carrying one SCD (>13 for \DMX, >40 for \TMX) prevents calculation in the correlated structure. However one can get the hyperfine constant between one electron spin of a \DMX elementary cell without exchange coupling with neighbors. The calculation was performed on a molecular unit cell made of 2 DMTTF molecules  and 8 surrounding iodine counter anions (see SI).     

We obtained for the g-factor: $g_{min}=2.0022$ and $g_{max}=2.0118$ in fair agreements with the experimental results (Fig. \ref{Fig:EPR1}). Note that g-factor is a local parameter which does not depend on the exchange parameter $J$. 
Then we focus on the hyperfine coupling between an electron spin and the nuclear spins. All the atoms present in (DMTTF)$_2$I have at least one isotope with nuclear spin.   
 
 \begin{figure}[h]
 	\centering
 	\includegraphics[width=0.49\textwidth]{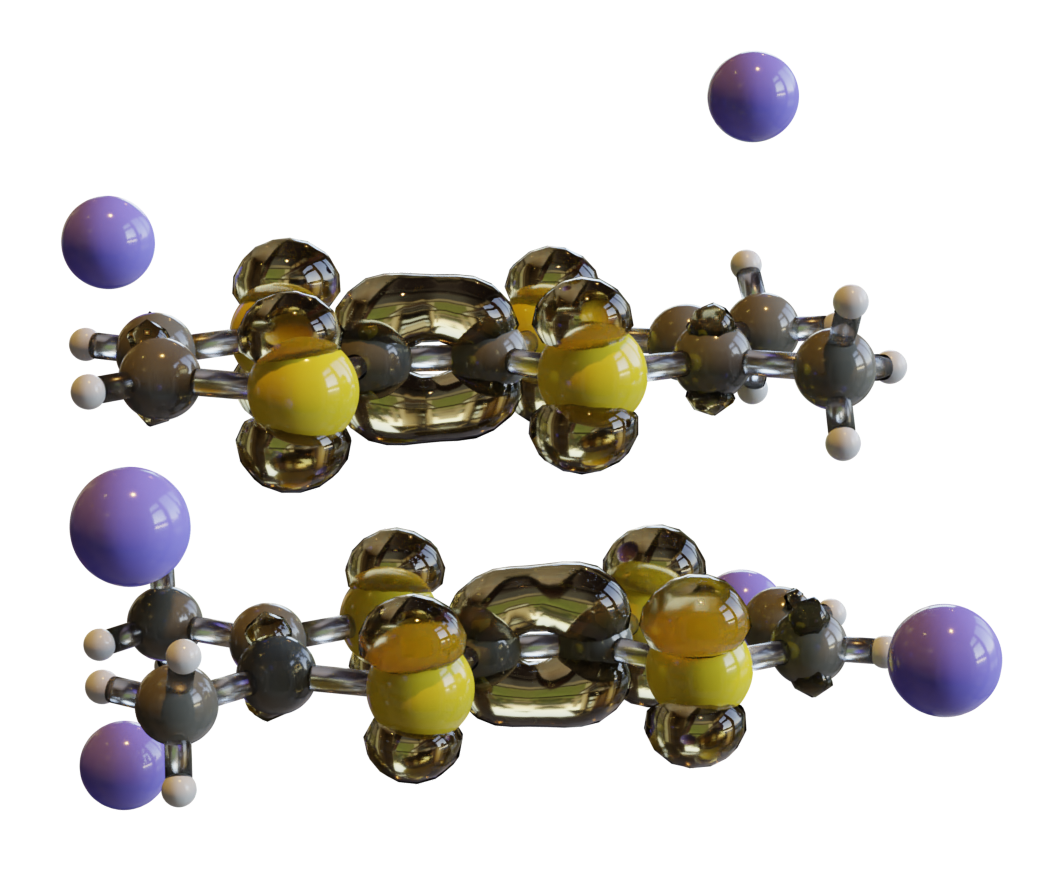}
 	\caption{Spin density plot of (DMTTF)$_2$I calculated by DFT. The color legend is : Carbon in black, Hydrogen in white, Sulfur in Yellow Iodine in blue. The glassy contour is the isodensity of spin. figure realized with Blender. }
 	\label{Fig:DFT}
 \end{figure}

Fig. \ref{Fig:DFT} shows the spin density distributed on the molecular cell obtained from the DFT calculation. We observe that the spin (total $S=1/2$) is split between the 2 DMTTF, and is distributed on several atoms, mostly on the central carbons and the sulfur atoms. In that plot (isosurface 0.02), the density of spin found  on the protons, outer carbons and iodines is very weak. As a consequence, the hyperfine coupling  expected for such a molecule "in absence of exchange coupling" should be large. Table 1 shows the mean values of hyperfine coupling for the nuclear spins present in the molecule. The full table for all sites is given in supplementary information.   
\begin{table}
\begin{tabular}	{|c|c|c|c|c|}
	\hline
	Center & $A_{min}$ & $A_{mid}$ & $A_{max}$ & $A_{iso}$ \\
	\hline
	$^{33}$S & -2.6 & -3.4 & 23.7 & 5.9 \\
	\hline
	$^{13}$C$_{iner}$ & -5 & -5.6 & 15.5 & 1 \\
	\hline
	$^{13}$C$_{outer}$ & -1.1 & -2.0 & 3.2 & 0.8 \\
	\hline
	$^{1}$H & 0.3 & 0.4 & 1.5 & 0.5 \\
	\hline
	$^{127}$I & 1.8 & 1.8 & 2.6 & 2 \\
	\hline
\end{tabular}
\label{Tab:coupling}\caption{Mean average values of the hyperfine coupling constant of the nuclear spin isotope (MHz).  }
\end{table}
The coupling constants calculated by DFT are much larger than the ones obtained from the 2D-HYSCORE. We attribute this large difference to the presence of exchange interaction between the molecules of \DMX. The SCD is made by about 13 spins for \DMX  and more than 40 for \TMX and the correlation between each local spin involved in the creation of the SCD acts like the exchange narrowing effect \cite{Anderson1953} which reduce all sources of second order moment of the EPR line. The data from DFT calculations combined to the experimental ones show an effect of screening of the hyperfine interaction by the exchange coupling.

\section{Conclusion}
Defects strongly correlated to a spin chain are $N-$body objects with unconventional properties. This theoretical study shows that the number of spins involved in the creation of the SCD directly depend on the dimerized property of the medium (here \DMX and \TMX). The energetic structure of such object is similar to the one found in other molecular magnet like V15. cw-EPR shows that the low temperature signal belongs to the SCD and not to an extrinsic impurity. Pulsed EPR shows that the SCD is coupled to the nuclear spins of the bath but the hyperfine interaction is strongly reduced (and almost negligible) compared to the  uncoupled molecule. 

\begin{acknowledgements}
EPR measurements at IM2NP and BIP Marseille as well as at LASIR Lille were supported by the Centre National de la Recherche Scientifique (CNRS) research infrastructure RENARD (Grant No. IR-RPE CNRS 3443). We thank G. Gerbaud for technical assistance.   
\end{acknowledgements}

\bibliography{InvitationDMTTF.bib}  

\end{document}